# A WIMP detector with two-phase liquid xenon[*]

David B. Cline

Department of Physics and Astronomy, Box 951647
University of California Los Angeles, Los Angeles, CA 90095-1647, USA

We describe the liquid-xenon dark-matter detector program of the UCLA–Torino team. A two-phase detector, ZEPLIN II, for the Boulby Mine is a good match for the current search for WIMP dark matter.

## 1. INTRODUCTION

The search for cold dark-matter WIMP particles is now entering the level of supersymmetric particle sensitivity. It is considered that the most likely neutralino mass is of order 100 GeV and the expected interaction rate is from $10^{-1} - 10^{-4}$ events/kg/day. In order to carry out a successful search, we need (1) a massive (~50kg) powerful background-discriminating detector, and (2) a high-A target to match the effective mass of the WIMP particles to give the largest recoil energies.

Over the past few years, the UCLA–Torino ICARUS team has been developing just such a detector with H. Wang (UCLA) and P. Picchi (Torino), who are the leading contributors to this research. In this paper we discuss the progress in this area and the plans to construct a 40-kg detector (ZEPLIN II) for the search for WIMP dark matter in the near future.

## 2. THE STATUS OF THE SEARCH FOR WIMP DARK MATTER

Figure 1 gives the current limits on the WIMP search, including the possible DAMA signal, that have been presented at this meeting. In essence, we have reached the level of about 1 event/kg/d as a limit or an observation (DAMA). Most theoretical models for neutralino interaction give a lower rate (i.e., $<10^{-1}$/event/kg/d) expectation, but the Bottino group at Torino has shown that the DAMA observations can be made consistent with the theory. In Fig. 1, we also show the projected sensitivity of a number of new detectors such as CDMS II and ZEPLIN II, which are discussed in this report.

## 3. THE DEVELOPMENT OF A TWO-PHASE DISCRIMINATING DETECTOR

Starting in the early 1990s, the UCLA–Torino ICARUS group initiated the study of liquid Xe as a WIMP detector with powerful discrimination. Our most recent effort is the development of the two-phase detector. Figure 2 shows our 1-kg, two-phase detector and the principle of its operation. WIMP interactions are clearly discriminated from all important background by the amount of free electrons that are drifted out of the detector into the gas phase where amplification occurs. In Fig. 3, we show the resulting separation between backgrounds and simulated WIMP interactions (by neutron interaction). It is obvious from this plot that the discrimination is very powerful.

## 4. THE DESIGN AND CONSTRUCTION OF ZEPLIN II

After the success of the 1-kg, two-phase detector, two directions are being taken: (1) Construction of a large two-phase detector to search for WIMPs. The UCLA–Torino group has formed a collaboration with the UK Dark Matter team to construct a 40-kg detector (ZEPLIN II) for the Boulby Mine underground

---

laboratory (Fig. 4). (2) Continuation of the R&D effort with liquid Xe to attempt to amplify the very weak WIMP signal. This work will constitute the Ph.D. thesis of J. Woo (UCLA). The first idea to test is inserting a CsI internal photo cathode to convert UV photons to electrons that are subsequently amplified by the gas phase of the detector.

## 5. ACKNOWLEDGMENTS

I wish to acknowledge the excellent work of H. Wang and P. Picchi, and of the whole ICARUS team as well. In addition, I have benefitted from interactions with P. F. Smith, N. Smith, and N. Spooner and the rest of the UK Dark Matter team.

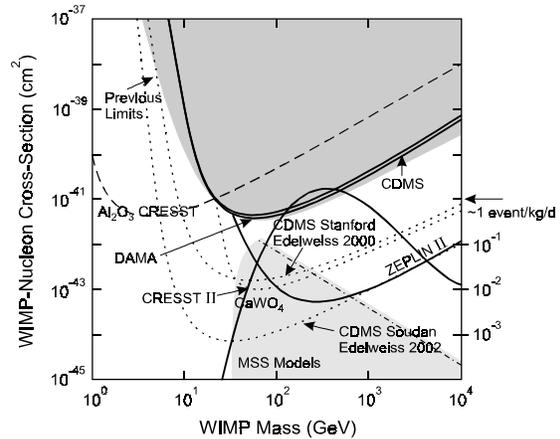

Figure 1. Current limits and prospects for a dark matter detector with ZEPLIN II.

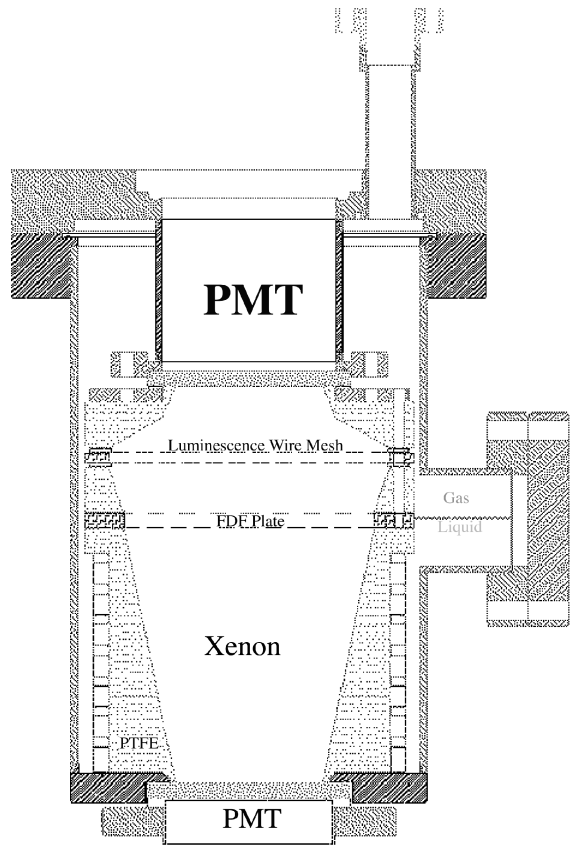

Figure 2. One-kg, two-phase Xe WIMP detector.

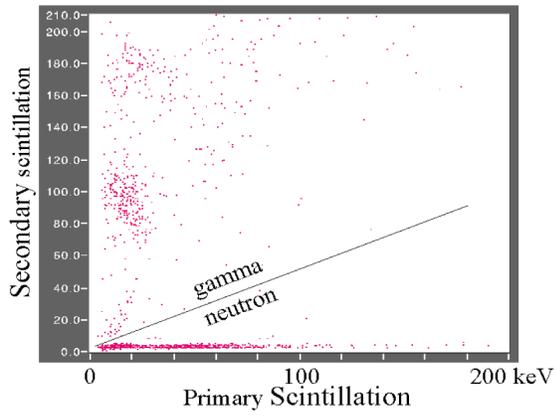

Figure 3. Secondary vs primary scintillation plot in two-phase Xe with mixed gamma and neutron sources. The secondary scintillation photons are produced by electroluminescent process in gaseous Xe.

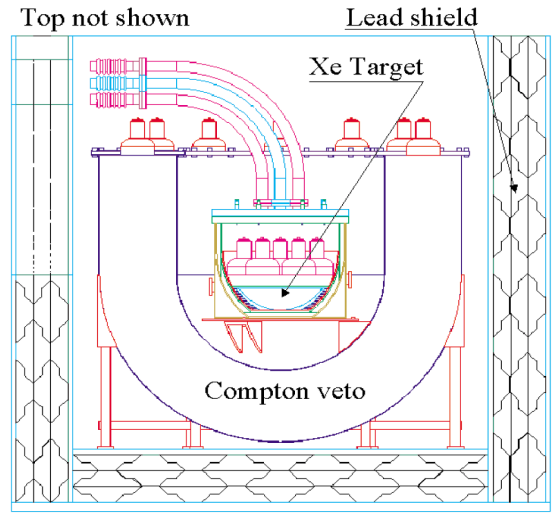

Figure 4. Conceptual design for the 30-kg ZEPLIN II detector proposed by the UCLA‑Torino‑UK group.